\documentstyle[prb, aps, twocolumn, floats]{revtex}

\newcommand{\half}[1]{ \frac{ #1 }{ 2 } }
\newcommand{\Tr}{ {\text{Tr}} }
\newcommand{\nl}{ \nonumber\\ }
\newcommand{\nid}{ \noindent }
\newcommand{\Ree}{ {\text{Re}} }
\newcommand{\Imm}{ {\text{Im}} }

\begin{document}
\input epsf.sty

\twocolumn[
\hsize\textwidth\columnwidth\hsize\csname@twocolumnfalse\endcsname

\draft

\title{Dynamical Behavior of the Dissipative Two-State System}
\author{Klaus V\"olker} \address{Department of Physics and Astronomy,
  University of California, Los Angeles, CA 90095} \date{\today}

\maketitle

\begin{abstract}
  We investigate the dynamical correlation function of a
  quantum-mechanical two-state system which is coupled to a bosonic
  heat bath, utilizing the equivalence between the spin-boson
  Hamiltonian and the $1/r^2$ Ising model. The imaginary-time
  correlation function is obtained from Monte Carlo simulations on the
  Ising system and then continued to real time by a Pad\'e
  approximation.  In the unbiased system, the transition from
  oscillatory to strongly damped behavior is found to occur at a
  coupling strength close to $\alpha = 1/2$.  The biased system favors
  coherent relaxation and displays a significantly larger crossover
  value $\alpha_c$. We introduce the quasiparticle picture to describe
  the relevant behavior at intermediate time scales.  Within this
  approximation, we map out phase diagrams for the unbiased and biased
  systems.
\end{abstract}



]

\section{Introduction}

The quantum mechanical two-state system, coupled to a dissipative
environment, provides a universal model for many physical systems. An
important example of current interest is the problem of
defect-tunneling in solids.\cite{GoldingCoppersmith} Our interest in
this problem was renewed when connections between current topics in
high temperature superconductivity \cite{ChAnderson} and quantum Hall
effect \cite{LesSaleur} were perceived. In fact, all systems that can
be described by a double-well potential associated with a generalized
coordinate, with appropriate restrictions on the parameters, reduce to
a two-state system at sufficiently low temperatures, when only the
ground states of the two individual wells remain relevant. \cite{RMP}

At low temperatures, we can choose the localized ground states of the
two potential wells, in the absence of tunneling, as the basis of our
two-dimensional Hilbert space. The overlap of the two wavefunctions
leads to quantum mechanical tunneling, described by a transition
matrix element between the two effective states of the system.
Decoupled from its environment, we can write the Hamiltonian of the
system in pseudospin language as

\begin{equation}
        H_{0} = - \half{\Delta} \sigma_x + \half{\epsilon} \sigma_z.
\end{equation}

\nid $\sigma_x$ and $\sigma_z$ are the usual Pauli matrices, so that
$\Delta/2$ is the tunneling matrix element, and $\epsilon$ describes
the bias of the system, i.e., the difference in the ground state
energies of the two localized states. The isolated system is trivial
to diagonalize and exhibits coherent tunneling between the two states.

If the system is coupled to a dissipative environment
\cite{WeissBook}, we can expect the tunneling to lose its phase
coherence. This can happen even at zero temperature if the continuous
spectrum of the macroscopic dissipative environment extends down to
zero frequency. A particularly elegant model of the environment has
been known for some time. \cite{RMP} In this model, a set of harmonic
oscillators are coupled linearly to $\sigma_z$. The full Hamiltonian
is

\begin{eqnarray}
\label{EqnTS}
        H &=& - \half{\Delta} \sigma_x + \half{\epsilon} \sigma_z +
                 \sum_{\alpha} \omega_\alpha a_\alpha^\dagger a_\alpha \nl
                 &+& \half{\sigma_z } \sum_{\alpha} { f_\alpha (
                 a_\alpha^\dagger + a_\alpha ) },
\end{eqnarray}

\nid where the $a_\alpha$ are destruction operators of the harmonic
oscillators with frequencies $\omega_\alpha$. The quantity $f_\alpha$
represents the coupling strength of the two-state system to the
coordinate of the $\alpha^{th}$ oscillator. This is also known as the
spin-boson Hamiltonian \cite{RMP} that was proposed in the context of
a model of a local magnetic impurity coupled by spin-flip scattering
to the conduction electrons of the host metal, known as the Kondo
problem. The low-energy particle-hole excitations of the conduction
electrons define the oscillator states. The environment is completely
characterized by its spectral density $J(\omega)$

\begin{equation}
        J(\omega) = \half{\pi} \sum_{\alpha} { f_\alpha^2 \,
                    \delta(\omega - \omega_\alpha) },
\end{equation}

\nid As a model of a linear dissipative environment, we consider an
Ohmic environment that classically exhibits friction of the form $F =
- \eta \dot{q}$, and is described by the spectral density \cite{RMP}:

\begin{equation}
\label{EqnSpec}
        J(\omega) = \Bigl\{ \begin{array}{c} 2\pi\alpha\omega, \; \;
                    \omega \ll \omega_c \\ 0, \; \; \omega \gg \omega_c
                    \end{array} \Bigr. .
\end{equation}

\nid The assumption of an Ohmic bath reduces the characteristics of
the environment to two parameters, the dimensionless coupling strength
$\alpha$ and the characteristic cutoff scale $\omega_c$. The
generality of this model is not obvious, unless the phenomena to be
studied are strongly dominated by low energy processes. Indeed, the
results of the present paper can be trusted for only such low energy
processes, which are akin to dynamic quantum critical phenomena. We
shall be interested in the dynamical correlation function

\begin{equation}
        C(t) = \frac{1} {2i} \left< \left[ \sigma_z(t), \sigma_z(0) \right]
               \right>,
\end{equation}

\nid or its Fourier transform, the response function $\chi''(\omega)$,
at zero temperature.  In particular, we will consider its qualitative
and quantitative dependencies on the parameters $\Delta$, $\epsilon$
and $\alpha$, appearing in the guise of dimensionless ratios
$\Delta/\omega_c$ and $\epsilon/\omega_c$ ($\hbar$ and $k_B$ are set
to unity).

In an analogous model, one can also couple a two-state system to a
Fermionic environment that can be described by a non-interacting set
of quasiparticles, \cite{ChangCh} or to a Fermi liquid. Except for
certain restrictions on parameters, this is equivalent to the
oscillator model for the low-energy states of the tunneling degree of
freedom. More intriguing, however, is the case where the tunneling
degree of freedom is coupled to a Luttinger liquid as in a 
quasi-one-dimensional electron gas, or to the edge states in a quantum Hall
system, \cite{LesSaleur} but this topic is outside the scope of the
present paper.

It is known for some time \cite{AndersonYuval} that the partition
function of this system can be transformed into the partition function
of an Ising model with long-range interactions. A particular Ising
site corresponds to the pseudospin state at a particular point on the
imaginary-time axis. This connection enables us to determine the
imaginary-time correlation function of the two-state system from Monte
Carlo simulations of the Ising model. A Wick rotation then leads us
from imaginary time to the above mentioned real-time correlation
function $C(t)$. This procedure will be described in detail in the
following sections and constitutes a reliable \emph{nonperturbative}
evaluation of the dynamics of the system in regimes in which dilute
blip approximation employed previously is not reliable. The present
paper is based on ideas introduced in a short communication by
Chakravarty and Rudnick. \cite{ChRudnick}

We know that the correlation function $C(t)$ will exhibit coherent
oscillations for zero coupling. Weak coupling to the environment leads
to damped oscillatory behavior. For small $\alpha$ and
$\Delta/\omega_c$, scaling arguments \cite{Guinea} yield the
renormalized tunneling frequency

\begin{equation}
  \Delta_r = \Delta \left( \Delta/\omega_c \right) ^{\alpha/1-\alpha}.
\end{equation}

\nid If the influence of the environment is strong enough, we expect a
completely incoherent decay of the time correlation function; further
increase of the coupling leads to broken symmetry and the tunneling
degree of freedom is localized. \cite{Chakravarty}  One aim of the
present paper is to map out the coherent and incoherent regimes in a
phase diagram, at zero temperature, and to provide quantitative
results for the behavior near the crossover value of $\alpha$. We will
also obtain a qualitative picture of the behavior of $C(t)$ in a
biased system $(\epsilon \ne 0)$.

In Sec.\ \ref{lblSection2} we will derive the precise correspondence
between the spin-boson Hamiltonian and a classical Ising spin system.
Section \ref{secNumerical} deals with aspects of the Monte Carlo
simulations on the Ising model, and with the Pad\'e approximation
method used to obtain real-time results. Those results are presented
in Sec.\ \ref{secResults}. Section \ref{secAnalytic} introduces the
quasiparticle picture as an elegant way to model the essential physics
at intermediate times. Results obtained within the quasiparticle
approximation are presented in Secs.\ \ref{secOscillatory} and
\ref{secBias}. The last section compares our results with some exact
results for the zero-frequency limit of the spectral function and for
the Toulouse case $\alpha = 1/2$.

\section{The dissipative two-state system, the Coulomb gas model 
  and the inverse-square Ising model}
\label{lblSection2}

The partition function corresponding to the Hamiltonian (\ref{EqnTS})
can be cast into the Coulomb-gas \cite{SchotteSchotte} form, that is,
a one-dimensional system of alternating positive and negative charges.
Expanding the partition function in terms of $\sigma_x$ in imaginary
time, we get:

\begin{eqnarray}
        Z &=& \Tr \Bigl[ e^{-\beta H} \Bigr] = \Tr \Bigl[ e^{-\beta
          H_{\infty} } \sum_{n=0}^\infty (-1)^n \nl
           & & \times \int_0^\beta d\tau_1 \cdots
            \int_0^{\tau_{n-1}} d\tau_n \: H_{\Delta}(\tau_1) \cdots
            H_{\Delta}(\tau_n) \Bigr].
\end{eqnarray}

\nid where

\begin{equation}
        H_{\Delta} = - \half{\Delta} \sigma_x, 
\end{equation}

\begin{equation}
        H_{\infty} = \sum_{\alpha}
        \omega_\alpha a_\alpha^\dagger a_\alpha + \half{\sigma_z}
        \left( \sum_{\alpha} { f_\alpha ( a_\alpha^\dagger + a_\alpha
        ) } + \epsilon \right),
\end{equation}

\nid and $H_{\Delta}(\tau) = e^{\tau H_{\infty}} H_{\Delta} e^{-\tau
  H_{\infty} }$ is the operator in the interaction representation.
After integrating out the environmental degrees of freedom, we arrive at
 the following expression for $Z$:

\begin{eqnarray}
\label{eqnPartFunc2}
        Z &=& Z_0 \sum_{n=0}^\infty \left( \half{\Delta} \right)^{2n}
            \int_0^\beta d\tau_1 \cdots \int_0^{\tau_{2n-1}}
            d\tau_{2n} \nl
          & & \times \exp \Bigl\{ {-\half{\epsilon} \int_0^\beta ds
            \: \xi(s) - \tilde{H} } \Bigr\},
\end{eqnarray}

\nid where

\begin{eqnarray}
        \tilde{H} &=& -\frac{1}{8} \sum_\alpha f_\alpha^2 \int_0^\beta
                    ds \: \xi(s) \int_0^\beta ds' \: \xi(s') \nl
                  & & \times \Bigl[
                    e^{-\omega_\alpha |s-s'|} + 2n_\alpha \cosh
                    \omega_\alpha(s-s') \Bigr],
\end{eqnarray}

\begin{eqnarray}
        \xi(s) &=& \Bigl\{ \begin{array}{c} +1, \; \; \tau_{2k} < s <
                 \tau_{2k+1} \\ -1, \; \; \tau_{2k-1} <
                 s < \tau_{2k} \end{array} \Bigr. \nl
               &=& 1 + 2 \sum_{n=0}^\infty (-1)^n \theta(s-\tau_n),
\end{eqnarray}

\nid The algebraic manipulations are detailed in Ref.\ 
\onlinecite{ChangCh}. $n_\alpha = 1/(e^{\beta \omega_\alpha} - 1)$ is the Bose
occupation factor, and $Z_0$ is the partition function of the
environment. At this point we will introduce the explicit form of the
spectral density:

\begin{equation}
        J(\omega) = \pi \sum_\alpha f_\alpha^2
                    \delta(\omega-\omega_\alpha) = 2 \pi \alpha \omega
                    \tilde{\theta}(\omega/\omega_c).
\end{equation}

\nid Here, $\tilde{\theta}(x)$ symbolizes a generic cutoff function: $
\tilde{ \theta }(x) = 1$ for $x = 0$, $\tilde{\theta}(x) = 0$ for $x
\gg 1$, and smooth in between. Then

\begin{eqnarray}
        \tilde{H} &=& -\frac{\alpha}{4} \int_0^\beta ds \: \xi(s)
                    \int_0^\beta ds' \: \xi(s') \nl
                  & & \times \int_0^\infty dw \: w
                    \frac{ \cosh \bigl[ \omega \bigl(
                    \half{\beta}-|s-s'| \bigr) \bigr] }{ \sinh
                    \half{\beta \omega} }
                    \tilde{\theta}(\omega/\omega_c).
\end{eqnarray}

\nid An approximate treatment of the cutoff consists in removing the
cutoff from the integral in this expression, and reintroducing it in
the form of the regularization condition $|\tau-\tau'| \ge
\omega_c^{-1}$ in the partition function (\ref{eqnPartFunc2}). This
corresponds to a hard-sphere repulsion between the Coulomb gas
charges. Then the integral over $\omega$ yields:

\begin{eqnarray}
        \int_0^\infty dw \: w \frac{ \cosh \bigl[ \omega \bigl(
        \half{\beta}-|s-s'| \bigr) \bigr] }{ \sinh \half{\beta \omega}
        } \nl
        = \left( \frac{\pi}{\beta} \right)^2 \Bigl[ \sin \left(
        \frac{\pi}{\beta}|s-s'| \right) \Bigr]^{-2}.
\end{eqnarray}

\nid Next we can perform a two-fold partial integration over $s$ and
$s'$, noting that the boundary terms are zero and $d\xi/ds = 2 \sum_k
(-1)^k \delta(s-\tau_k)$, arriving at:

\begin{eqnarray}
\label{EqnCoulombGas}
        Z &=& Z_0 \sum_{n=0}^\infty \left( \half{\Delta \tau_c}
              \right)^{2n} \int_0^\beta \frac{d\tau_1}{\tau_c}
              \int_0^{\tau_1-\tau_c} \frac{d\tau_2}{\tau_c} \cdots \nl
          &\times& \int_0^{\tau_{2n-1} - \tau_c} \frac{d\tau_{2n}}{\tau_c}
              \exp \biggl\{ \epsilon \sum_{j=1}^{2n} (-1)^j
              \tau_j \nl
          & & + 2 \alpha \sum_{i<j} (-1)^{i+j} \ln \left|
              \frac{\beta}{\pi\tau_c} \sin \frac{\pi}{\beta} (\tau_j -
              \tau_i) \right| \biggr\}
\end{eqnarray}

\nid The $\tau_c = 1/\omega_c$ appearing in the upper boundaries of
integration is the regularization condition that replaced the
high-frequency cutoff in $J(\omega)$. In this picture, the positions
$\tau_j$ of the charged particles correspond to spin-flips of the
two-state system on the imaginary-time axis. It should be emphasized
that this treatment of the cutoff is somewhat artificial, and we
therefore do not have a complete description of the high-energy
details of the system. Note, however, that the logarithm in the above
expression (\ref{EqnCoulombGas}) vanishes as $\tau - \tau' \to
\tau_c$, so that our treatment is self-consistent. A more rigorous
treatment would lead to additional terms of the form $(\tau_j -
\tau_i)^{-x}, x \ge 1$ in the Hamiltonian. The effects of these terms
on the results for intermediate time scales can be summarized by
replacing $\Delta \tau_c$ with an effective value $(\Delta
\tau_c)_{\text{eff}}$. Otherwise, the cutoff procedure does not
affect our results for intermediate time scales. \cite{AndYuvHam}
Furthermore, since the $ln$ is the only scale invariant term in the
series, the additional terms cannot change the critical behavior of
the system. \cite{Cardy}

Let us now consider a one-dimensional Ising model with long-range
interactions and periodic boundary conditions, given by:

\begin{equation}
        Z = \sum_{S_1 \cdots S_N} \exp \Bigl\{ - \sum_{j<i} V(i-j) S_i
            S_j - h \sum_j S_j \Bigr\}.
\end{equation}

\nid Following Cardy, \cite{Cardy} we can rewrite the partition
function in terms of interactions between spin flips or kinks:

\begin{eqnarray}
        Z &=& \sum_{n=0}^\infty y^{2n} \int_0^\beta \frac{d\tau_1}{a}
            \int_0^{\tau_1-a} \frac{d\tau_2}{a} \cdots
            \int_0^{\tau_{2n-1}-a} \frac{d\tau_{2n}}{a} \nl
          &\times&
            \exp \Bigl\{ \sum_{j<i} (-1)^{i-j} 4 U\left(
            \frac{\tau_i-\tau_j}{a} \right) \nl 
          & & + 2h \sum_j (-1)^j \left(
            \frac{\tau_j}{a} \right) \Bigr\},
\end{eqnarray}

\nid where $U(k)$ is defined by

\begin{equation}
\label{EqnUdef}
        V(x) = U(k+1) -2U(k) + U(k-1), 
\end{equation}

\nid and $y = e^{2U(0)}$ is the chemical potential, or fugacity, of
the system.

The short-distance cutoff at $\tau = a$ was introduced to recapture
the high-energy properties of the discrete lattice, after allowing the
kinks to move in the continuous interval $\tau = 0 \cdots Na$. Here,
$a$ is the lattice constant of the Ising model. Again, this
approximate treatment of the cutoff changes the high-energy behavior
of the system, but leaves the low-energy physics invariant, aside from
a modification of parameters related to the cutoff.

Comparison with (\ref{EqnCoulombGas}) yields $a = \tau_c$ and

\begin{equation}
        U(n) = \half{\alpha} \ln \left| \frac{N}{\pi} \sin \frac{\pi
               n}{N} \right|, \; \; n \ge 1,
\end{equation}

\nid so that

\begin{equation}
        V(n) = - \half{\alpha} \frac{ (\pi/N)^2 }{ \sin^2 (\pi n/N) },
               \; \; n \ge 2,
\end{equation}

\nid where we neglected terms of order $(\pi/N)^4$. $V(1)$ is
determined by $U(0)$, which in turn depends on the fugacity $y =
e^{2U(0)} = \Delta\tau_c/2$.  In the limit $N \to \infty$ we get:

\begin{equation}
        U(0) = V(1) - \half{\alpha}\gamma,
\end{equation}

\nid where $\gamma = 0.577\ldots$ is Euler's constant. In the familiar
language of the Ising model, the Hamiltonian reads:

\begin{eqnarray}
\label{IsingZ}
        \beta_I H_I = &-& \half{J_{NN}} \sum_i S_i S_{i+1} \nl
        &-& \half{J_{LR}} \sum_{j<i} \frac{ (\pi/N)^2 S_i
                      S_j }{\sin^2[\pi(j-i)/N] } 
                    - h \sum_i S_i,
\end{eqnarray}

\nid where $J_{LR} = \alpha$, and $J_{NN} + J_{LR} = - 2 V(1)$, so
that $y = e^{-J_{NN} - (1+\gamma) J_{LR}}$. To summarize, the
correspondence of the parameters is as follows:

\begin{eqnarray}
        \frac{\Delta_{\mathit{eff}}}{2\omega_c} &=& \exp \lbrace
             {-J_{NN}-(1+\gamma)J_{LR}} \rbrace; \nl
        \alpha &=& J_{LR}; \nl
        \frac{\epsilon}{2\omega_c} &=& h; \nl \beta \omega_c &=& N.
\end{eqnarray}

\nid Finally, the anisotropic spin-half Kondo model can be cast into
the same Coulomb gas form. \cite{AndersonYuval} In particular, the
case $\alpha = 1/2$ corresponds to the exactly solvable Toulouse limit
of the Kondo problem, which constitutes an important check of our
numerical results.

\section{Numerical techniques: Monte Carlo simulation 
  and Pad\'e approximation}
\label{secNumerical}
          
Due to the equivalence of both models, a calculation of the spin-spin
correlation function $S_I( |i-j| )$ of the Ising model (\ref{IsingZ})
provides us with the imaginary-time correlation function ${\cal
  C}(\tau) = \left< \sigma_z(\tau) \sigma_z(0) \right> $ of the
two-state system. The former can be computed using standard Monte
Carlo techniques. Since the regions of parameter space of primary
interest to us lie in the vicinity of the Ising model phase
transition, the accuracy of standard Monte Carlo algorithms is greatly
suppressed by the phenomenon of critical slowing down: The
autocorrelation function, which measures the number of Monte Carlo
steps necessary to obtain two statistically independent
configurations, diverges at the phase boundary.  To get a correct
picture of the thermodynamics, an extremely high number of lattice
sweeps becomes necessary.

To circumvent this problem, we use the Swendsen-Wang algorithm,
\cite{SwendsenWang} where the Ising model is mapped onto a
percolation model: In every Monte Carlo step, bonds are drawn with
probability $P_{ij} = 1 - e^{-J_{ij}}$ between each two parallel
spins, and no bonds are drawn ($P_{ij} = 0$) if the spins are
antiparallel. The spins do not necessarily have to be nearest
neighbors, and $J_{ij}$ is the sum of their nearest-neighbor and
long-range interactions.  Then, spins which are directly or indirectly
connected by bonds are grouped into clusters, and each spin in a
cluster is assigned the same spin state with a probability
$P_{\uparrow\downarrow} = e^{\pm nh}/(e^{nh}+e^{-nh})$ for spin-up and
spin-down, respectively. Here $n$ is the number of spins contained in
this particular cluster. It is straightforward to check that the
principle of detailed balance is satisfied for this algorithm, so that
the distribution of spin states converges to the Boltzmann
distribution. Our implementation of the algorithm performs the
formation of clusters simultaneously with the assignment of bonds, so
that no additional computation time has to be invested here.

Next, the Fourier transform ${\cal C}(\omega_n)$ at the Matsubara
frequencies is obtained. The Pad\'e approximant method of Vidberg and
Serene \cite{VidSerene} is used to continue this spectral function
from the positive Matsubara frequencies onto the real axis.
Essentially, the spectral function is approximated by a rational
function, with numerator and denominator polynomials of order $N/2$,
where $N$ is the number of Matsubara points used in the approximation.
This rational function can then be trivially continued into the
complex plane, given that the first quadrant of the plane is free of
singularities. The imaginary part of the resulting function is the
response function $\chi''(\omega)$, as we shall see below.

The Pad\'e approximation, in the context of analytic continuation, is
an ill-defined procedure in the sense that small errors in the input
data can lead to severe discrepancies in the output data.  Accuracy of
the simulation data is therefore a critical issue. We used $8 \times
10^6$ Monte Carlo sweeps on a lattice with $N = 256$ Ising spins for
most points in parameter space, but had to increase this number in the
vicinity of the crossover value of $\alpha$ to $2 \times 10^7$ sweeps.
The results are essentially independent of $N = \beta \omega_c$, so
they converge well in the zero-temperature limit.  However,
simulations on a finite spin system will not allow us to capture the
localization effects \cite{Chakravarty} that occur for $\alpha \to 1$.

To get a measure for the simulation errors, we calculated the Pad\'e
approximant several times for each data point, using different numbers
of lattice sweeps and Pad\'e points. The results were filtered by
checking the basic analytic structure of $\chi''(\omega)$ and the
behavior for $|\omega| \rightarrow \infty$.


 \begin{figure}
   \centerline{\epsfxsize=3.4in \epsffile{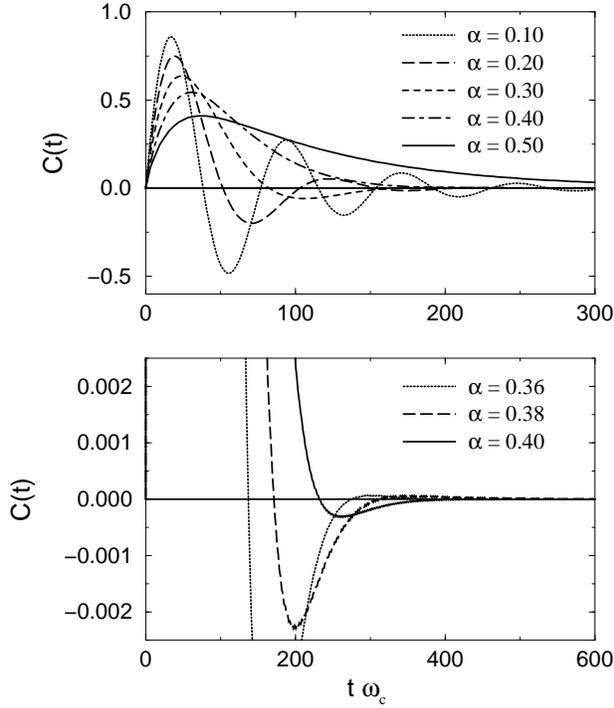}}
 \caption{The correlation function $C(t) = \frac{1}{2i}
   \left< \left[ \sigma_z(t), \sigma_z(0) \right] \right>$ for
   $\Delta/\omega_c = 0.1$ and different values of $\alpha$ (top).
   Oscillatory behavior can be observed up to a coupling strength
   $\alpha = 0.4$ (bottom).}
 \label{figTimeGraph}
 \end{figure}

\section{Results}
\label{secResults}

In this section we present results for the anti-sym\-metrized
correlation function $C(t) = \frac{1}{2i} \left< \left[ \sigma_z(t),
    \sigma_z(0) \right] \right>$ at zero temperature, as obtained from
Monte Carlo simulations on the Ising system (\ref{IsingZ}), and
continued to real time by a Pad\'e approximation.  The analytic
continuation is carried out in the complex frequency plane. Therefore
we introduce Green's functions corresponding to the correlation
functions in imaginary and real time as:

\begin{eqnarray}
        {\cal G}(\tau) &=& -\frac{1}{Z} \: \Tr \Bigl[  e^{-\beta H}
        e^{|\tau| H} \sigma_z e^{-|\tau| H} \sigma_z \Bigr] = - {\cal
        C}(|\tau|), \nl 
        G(t) &=&
        -\frac{i}{Z} \: \Tr \Bigl[ e^{-\beta H} e^{i H |t|} \sigma_z
        e^{-i H |t|} \sigma_z \Bigr] \nl 
        &=& -i \left< \sigma_z(|t|)\sigma_z(0)
        \right>.
\end{eqnarray}

\nid These Green's functions are bosonic in character (${\cal G}(\tau
+ \beta) = {\cal G}(\tau), \tau < 0$), so that their Fourier
transforms are connected in the complex plane as \cite{AGD}

\begin{equation}
        {\cal G}(-i \omega_n) = \Ree \, G(\omega) + i \: \Imm \,
        G(\omega) \tanh{ \half{\beta \omega} },
\end{equation}

\nid which leads to the following relation between ${\cal C}(\tau)$
and the response function $\chi''(\omega)$:

\begin{eqnarray}
\label{eqnWick}
        \chi''(\omega) &=& \half{1} \int_{-\infty}^\infty dt \: e^{i\omega
          t} \langle \left[ \sigma_z(t), \sigma_z(0) \right] \rangle \nl
       &=& \tanh{ \left( \frac{\beta\omega}{2} \right) } \: 
           \int_{-\infty}^\infty dt \: e^{i\omega t} 
           \Ree \langle \sigma_z(t) \sigma_z(0) \rangle \nl
       &=& \Imm \, {\cal C}(- i \omega).
\end{eqnarray}


\begin{figure}
  \centerline{\epsfxsize=3.4in \epsffile{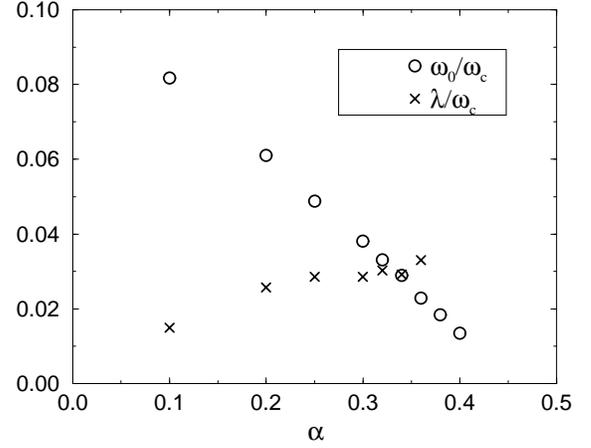}}
\caption{Oscillation frequency $\omega_0$ and damping coefficient
  $\lambda$ corresponding to the curves shown in Fig.\ 
  \ref{figTimeGraph}.}
\label{figTimeResults}
\end{figure}

Figure \ref{figTimeGraph} shows some examples \cite{CommentOnCt} for
$C(t)$ in the unbiased case ($\epsilon = 0$), for $\Delta/\omega_c =
0.1$ and various values of $\alpha$. We can clearly observe coherent,
weakly damped oscillations of the form $C(t) = Z \sin(\omega_0 t) \:
e^{-\lambda |t|}$ for small values of $\alpha$. A more detailed plot
(Fig.\ \ref{figTimeGraph} bottom) reveals that oscillations are indeed
visible up to $\alpha \simeq 0.4$. Past that value, the accuracy of our
numerical data does not allow to resolve oscillatory behavior anymore.
Figure \ref{figTimeResults} shows oscillation frequency $\omega_0$ and
damping coefficient $\lambda$ as obtained from the data in Fig.\ 
\ref{figTimeGraph}.

\begin{figure}
  \centerline{\epsfxsize=3.4in \epsffile{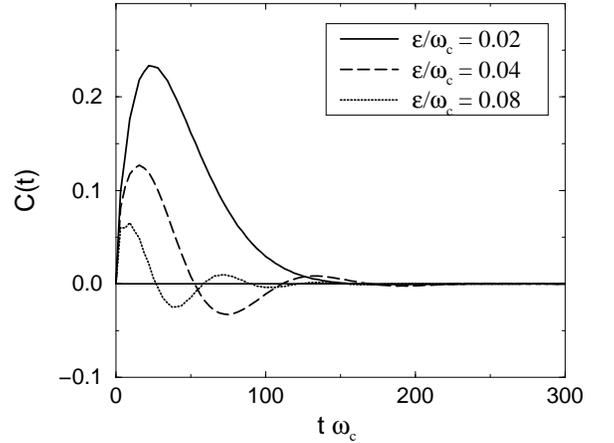}}
\caption{ $C(t)$ for the biased system, for $\Delta/\omega_c = 0.1, 
  \alpha = 0.5$ and various values of $\epsilon/\omega_c$. The biased
  system favors coherent oscillations.}
\label{figTimeBias}
\end{figure}

In the case of a biased system ($\epsilon \ne 0$, see Fig.\ 
\ref{figTimeBias}), the oscillation frequency is greatly enhanced,
while the damping coefficient remains largely unaffected by the bias.
The oscillations are weaker, since the system dwells in the
energetically favored state most of the time.


\section{Analytic Properties of the Spectral Function}
\label{secAnalytic}

\begin{figure}
  \centerline{\epsfxsize=3.4in \epsffile{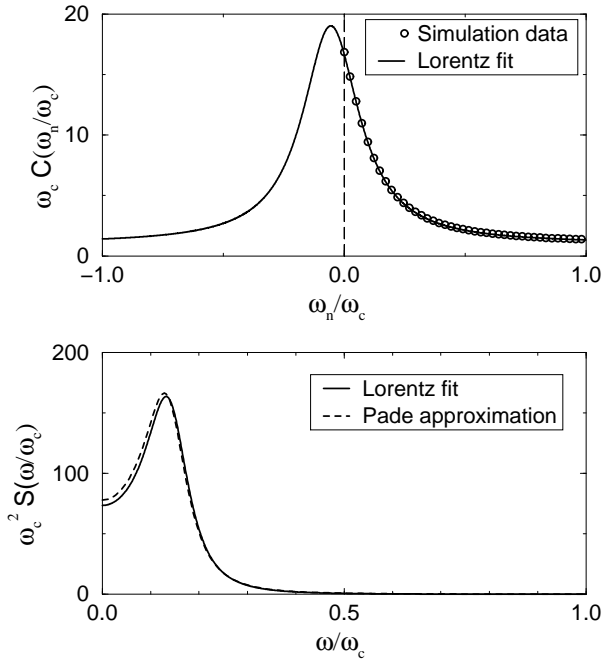}}
\caption{top: The imaginary-time correlation function $C(\omega_n)$ at
  the Matsubara frequencies, as obtained from Monte Carlo simulation
  (circles), for $\Delta/\omega_c = 0.2$, $\alpha = 0.2$, and its
  approximation by a Lorentzian (solid line).  Bottom: Spectral
  function $S(\omega)$, as calculated from the Lorentz fit (solid
  line), and by Pad\'e approximation (dashed line). }
\label{Lorentz}
\end{figure}

An example for the Fourier transform of the imaginary-time correlation
function ${\cal C}(\omega_n) = \int_0^\beta d\tau \: e^{i \omega_n
  \tau} \: {\cal C}(\tau)$ is shown in Fig.\ \ref{Lorentz}. Note that,
since ${\cal C}(\tau)$ is real and symmetric in $\tau$, only the
positive-frequency part has physical meaning. Analytic continuation to
negative frequencies reveals, however, that this function is very well
described by a Lorentzian:

\begin{equation}
        {\cal C}(\omega_n) \approx \frac{a w_0^3} { \left( \omega_n +
        \lambda \right)^2 + \omega_0^2 } + \mbox{const}.
\end{equation}


\nid This approximate form seems to hold extremely well in large parts
of parameter space. Typical deviations are of the order of 1\%, but
become larger close to $\alpha = 1/2$. However, we cannot expect this
simple approximation to give reliable results near the crossover value
of $\alpha$, and it will not be used in the following to obtain any
quantitative results. In particular,
it fails to reproduce the explicit form (\ref{eqnToulouse}) of the
spectral function for the Toulouse case $\alpha = 1/2$.

We consider the spectral function $S(\omega) =
\chi''(\omega)/{\omega}$. According to (\ref{eqnWick}), the
approximate form of $S(\omega)$ displays a simple four-pole structure
in the complex plane:

\begin{equation}
\label{eqnSfromLorentzian}
        S(\omega) \equiv \frac{\chi''(\omega)}{\omega} \approx 
        \frac{a Q^{-1}}{ \left[ (\omega/\omega_0)^2 -
        1 + Q^{-2} \right]^2 + 4 Q^{-2} }.
\end{equation}

\nid Here, $Q = \omega_0 / \lambda$ is the Q-value of the resonance.
$\omega_0$ is the oscillation frequency, and $\lambda$ the damping
coefficient. Again, this simplified form of the spectral function
cannot be expected to give reliable quantitative results and will be
used solely for illustrative purposes. All results are obtained from
Pad\'e approximants to ${\cal C}(\omega_n)$ based on up to 256
Matsubara points.

\subsection*{The Quasiparticle Picture}

\begin{figure}
  \centerline{\epsfxsize=3.4in \epsffile{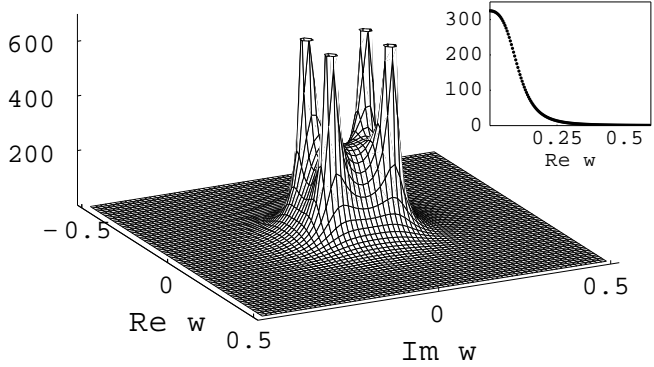}}
  \centerline{\epsfxsize=3.4in \epsffile{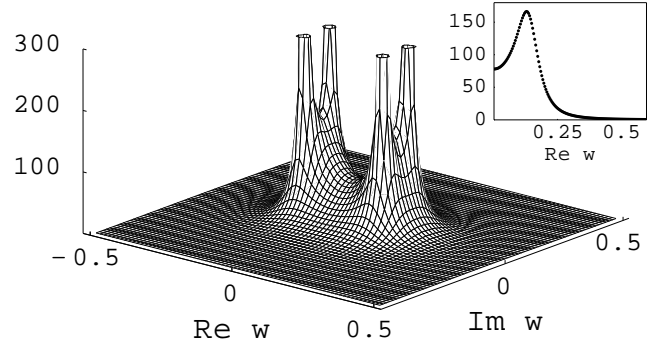}}
\caption{The absolute value of the spectral function $|S(\omega)|$ 
  in the complex $\omega$ plane, as obtained from a Pad\'e approximant
  using 256 Matsubara points, for $\Delta/\omega_c = 0.2$, and $\alpha
  = 0.2$ (top), $\alpha = 0.35$ (bottom). The analytic structure is
  dominated by four poles at $\omega = \pm \omega_0 \pm i \lambda$.
  The insets show $S(\omega)$ on the positive real axis: the
  quasiparticle peak is visible only for $\alpha < 0.33$. }
\label{PadePlane}
\end{figure}

Applying the Pad\'e approximation to obtain the precise form of
$S(\omega)$ in the complex plane, we discover that its analytic
structure is very similar to that of the Lorentzian approximation
(\ref{eqnSfromLorentzian}). It is dominated by four poles at $\omega =
\pm \omega_0 \pm i \lambda$, as can be seen in Fig.\ \ref{PadePlane}.
The physical meaning of these poles requires some consideration: We
know from the spectral representation of the correlation function that
it has a branch cut along the real axis, and is otherwise free of
singularities on the first Riemann sheet. If the spectral density
associated with the branch cut is sharply peaked around a finite
frequency $\omega_0$, it is legitimate to approximate the dominant
part of the spectral density by a simple pole at $\omega_0 + i
\lambda$, the ``quasiparticle pole'', which gives rise to the
oscillatory part of the correlation function:

\begin{equation}
  C(t) = Z \sin(\omega_0 t) \: e^{-\lambda |t|} + \mbox{(incoherent part)}
\end{equation}



\nid Alternatively, we can view the quasiparticle pole as resulting
from an analytic continuation of the spectral function past the branch
cut onto the second Riemann sheet. This is done, for example, in 
Fermi-liquid theory, and accurately reproduces the intermediate-time
behavior of the correlation function.  The intermediate-time range,
over which Fermi-liquid theory is valid, is defined by
\cite{FetterWalecka} $1/\epsilon_k \ll t \lesssim 1/\gamma_k$, where
$\epsilon_k$ and $\gamma_k$ are quasiparticle energy and damping,
respectively. In particular, neither the short-time decay proportional
to $t$ nor the exponential long-time decay implied by Fermi-liquid
theory reflect the true behavior of the correlation function.

In fact, the Pad\'e approximation naturally yields this analytic
continuation past the branch cut, since the existence of a branch cut
on the real axis cannot be inferred from the shape of $C(\omega)$ on
the imaginary axis.  The Pad\'e approximant therefore produces
precisely this ``quasiparticle picture'' of the spectral function.
The fact that it does not reveal any other characteristics indicates
that almost all of the spectral weight contributes to the
quasiparticle peaks. As in Fermi-liquid theory, the quasiparticle
picture will reproduce neither the high-energy behavior nor the
long-time tail of the correlation function. These are dominated by the
incoherent part of the spectral function, which corresponds to other,
more complicated singularities in the complex plane.

We might also consider the \textit{symmetrized} correlation function

\begin{equation}
C_s(t) = \Ree{ \left< \sigma_z(t) \sigma_z(0) \right> } \nl
       = \case{1}{2} \left< \sigma_z(t) \sigma_z(0) 
       + \sigma_z(0) \sigma_z(t) \right>,
\end{equation}

\nid which acquires an additional branch cut contribution (at T=0)
from the $tanh$ factor in Eqn.\ (\ref{eqnWick}):

\begin{equation}
\label{eqnIncoherent}
  C_s^{\text{inc}}(t) = \int_0^\infty \frac{d\omega}{2\pi} \: e^{-\omega t} 
  \left[ {\cal C}(\omega) - {\cal C}(-\omega) \right].
\end{equation}

Here, ${\cal C}(-\omega)$ is the analytic continuation of ${\cal C}(\omega)$
to negative frequencies. In particular, this term
leads to an asymptotic $1/t^2$ decay. 
When $Q \lesssim 1$, this incoherent background 
contribution will
become dominant at intermediate and short time scales as well, so that
in this regime it might be difficult to observe coherent oscillatory
behavior experimentally in the symmetrized correlation function
$C_s(t)$. As in Fermi-liquid theory, the time interval over which the
quasiparticle picture \textit{accurately} describes $C_s(t)$, is
bounded by $1/\omega_0 \ll t \lesssim 1/\lambda$, which implies $Q \gg
1$.

Note that these considerations do not apply to the anti-symmetrized
correlation function $C(t)$, since the contribution (\ref{eqnIncoherent})
is purely real. For this function, we expect deviations from the
quasiparticle behavior only at short-time scales comparable to the
inverse cutoff $1/\omega_c$. Furthermore, we have seen in Sec.\ 
\ref{secResults} that damped oscillatory behavior is observable at
time scales up to a few times the inverse damping coefficient. At
longer time scales, a crossover to algebraic decay may also be
observed in the anti-symmetrized correlation function $C(t)$.


\begin{figure}
  \centerline{\epsfxsize=3.4in \epsffile{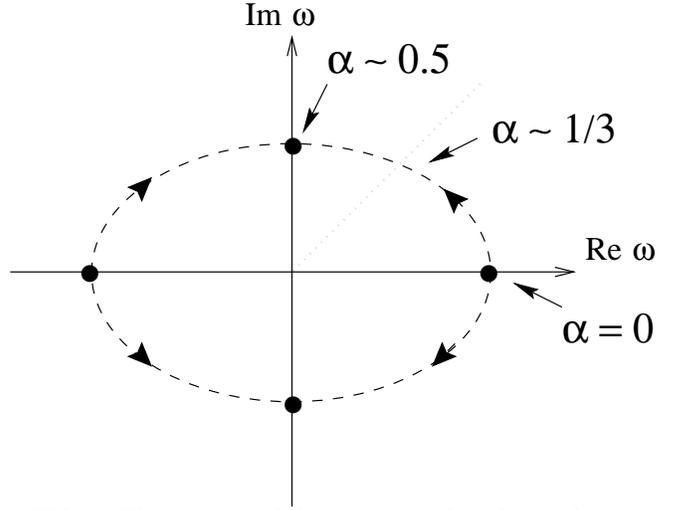}}
\caption{The journey of the quasiparticle poles in the 
  complex $\omega$-plane}
\label{Journey}
\end{figure}

\section{Oscillatory and damped behavior}
\label{secOscillatory}

The results presented here are based on the quasiparticle picture
introduced in the previous section. We have seen in Sec.\ 
\ref{secResults} that the quasiparticle picture offers an accurate
description of the intermediate-time dynamics at least for coupling
strengths up to $\alpha \approx 0.4$. Close to $\alpha = 1/2$, the 
spectral function may acquire a significant incoherent part due to
more complicated singularities in the complex plane, and the 
quasiparticle picture might not offer an accurate description of the
dynamics anymore. Therefore, the results very close to $\alpha = 1/2$
should best be viewed as characteristics of the quasiparticle model.

For $\alpha = 0$ (no coupling to the environment), each two
quasiparticle poles coincide on the real axis, yielding the familiar
oscillatory behavior of every quantum-mechanical two-state system. As
we turn on the coupling ($\alpha > 0$), the poles move away from the
real axis into the complex regime, as sketched in Fig.\ \ref{Journey},
corresponding to weakly damped oscillations. For even larger values of
$\alpha$, the poles meet with the imaginary axis, giving rise to
overdamped relaxation.

\begin{figure}
  \centerline{\epsfxsize=3.4in \epsffile{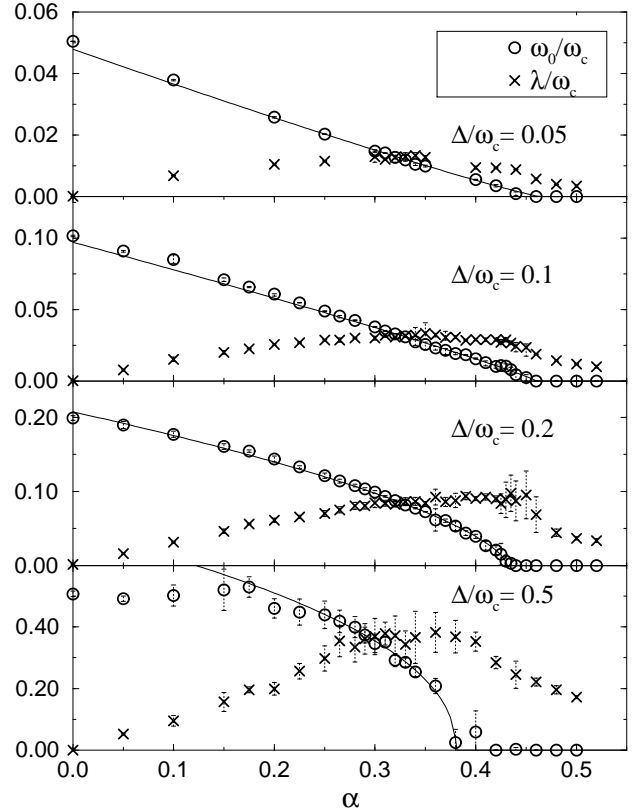}}
\caption{Tunneling frequency $\omega_0$ (circles) and damping
  coefficient $\lambda$ (X's) as functions of the coupling strength
  $\alpha$, for $\Delta/\omega_c = 0.05, 0.1, 0.2$ and $0.5$,
  respectively.  $\omega_0$ is strictly zero where no error bars are
  shown. The solid lines are power-law fits.}
\label{Results}
\end{figure}

The location of the poles in the complex plane was obtained
numerically from the results of the Pad\'e approximation. Fig.\ 
\ref{Results} shows the real ($\omega_0$) and imaginary ($\lambda$)
part of their location as a function of $\alpha$ for different values
of $\Delta/\omega_c$.  Note that the results agree very well with
Sec.\ \ref{secResults}.  The real part $\omega_0$ fits nicely to a
power-law curve: $\omega_0(\alpha) \sim (\alpha_c - \alpha)^\nu, \:
\alpha < \alpha_c$. Since we cannot {\it{ad hoc}} assume that the power-law
form holds for arbitrary $\alpha < \alpha_c$, only points with $\alpha
\ge 0.3$ were included in the fits. With the exception of
$\Delta/\omega_c = 0.5$, the graphs follow the power-law form
surprisingly well for all $\alpha < \alpha_c$. The results do not
change significantly if we limit ourselves to an even smaller number
of data points in the vicinity of $\alpha_c$.


\begin{figure}
  \centerline{\epsfxsize=3.4in \epsffile{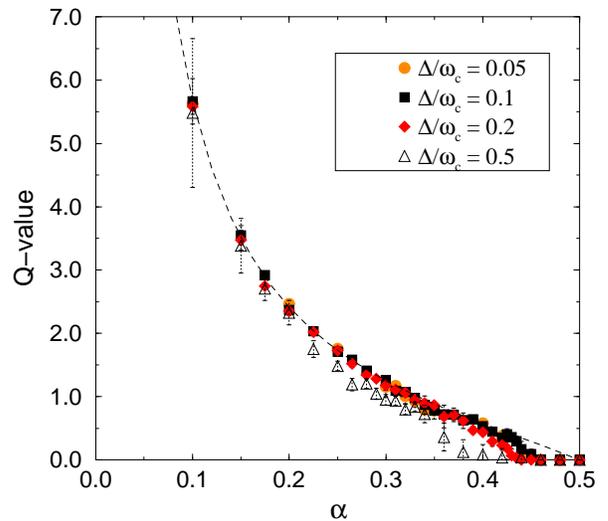}}
\caption{The $Q$ value $Q = \omega_0/\lambda$ as a function of $\alpha$, 
  for $\Delta/\omega_c = 0.05, 0.1, 0.2$ and $0.5$, respectively.  For
  $\Delta/\omega_c \le 0.2$ and $\alpha$ not too close to $\alpha_c$,
  the data points fall onto a universal scaling curve. The dashed line
  is expression (\ref{lblQforP}).}
\label{QValue}
\end{figure}

The $Q$ value, $Q=\omega_0/\lambda$, is shown in Fig.\ \ref{QValue}.
The scaling form \cite{Guinea}

\begin{equation}
  S(\omega) = \frac{1}{\Delta_r^2} f(\omega/\Delta_r, \alpha), 
\end{equation}

\nid which holds for $\Delta \ll \omega_c$, implies that the $Q$ value
is a universal function of $\alpha$, which is independent of
$\Delta/\omega_c$. This is indeed the case for $\Delta/\omega_c
\lesssim 0.2$ and $\alpha$ not too close to the critical value.

For values of $\alpha$ that are not too large, the oscillatory
behavior of the system is visible in the spectral function in the form
of inelastic peaks on the real axis, centered around $\omega = \pm
\omega_0$. If we increase the damping, the peaks broaden and move
closer to the origin. At $Q=1$, when the half-width of the peaks
starts to exceed their separation, they can no longer be
distinguished, and the spectral function now appears to be centered at
$\omega = 0$, as can easily be verified from the approximate form
(\ref{eqnSfromLorentzian}). The $Q$ factor reaches unity when $\alpha
\simeq 1/3$, in agreement with results obtained by a numerical
renormalization group calculation \cite{CostiKieffer} and an analytic
form factor approach \cite{LesSaleur}. Oscillatory behavior persists
beyond that point, however, due to the presence of the quasiparticle
poles in the complex plane at nonzero values of $\omega_0$. This was
confirmed in Sec.\ \ref{secResults} without resorting to the
quasiparticle picture. The physical meaning of $\alpha = 1/3$ is
simply that the damping coefficient starts to exceed the oscillation
frequency.

Within the quasiparticle point of view the transition to strongly
damped behavior takes place when the poles coincide with the imaginary
axis $(\omega_0 = 0)$.  For small $\Delta/\omega_c$, the equivalence
between the two-state system with $\alpha = 1/2$ and the Toulouse
limit of the Kondo model provides good reasons to assume that this
transition occurs at $\alpha_c = 1/2$.  Calculations of the
correlation function $P(t)$, which is the conditional average $\left<
  \sigma_z(t) \right>$ with the constraint $\sigma_z(t) = 1$ for $t <
0$, support this conclusion. \cite{RMP} An exact expression
\cite{SaleurNew} for the $Q$ value associated with $P(t)$ is known
in the limit $\Delta/\omega_c \to 0$:

\begin{equation}
\label{lblQforP}
  Q = \cot \left( \frac{\pi}{2} \frac{\alpha}{1-\alpha} \right).
\end{equation}

\nid This expression matches the data in Fig.\ \ref{QValue} almost
exactly, for $\alpha$ not too close to $1/2$. This is an indication
that $P(t)$ and $C(t)$ possess an equivalent structure at intermediate
times.  The discrepancy near $\alpha = 1/2$ may either indicate a
breakdown of the quasiparticle picture, or may arise from the fact
that we consider finite values of $\Delta/\omega_c$.


\begin{figure}
  \centerline{\epsfxsize=3.4in \epsffile{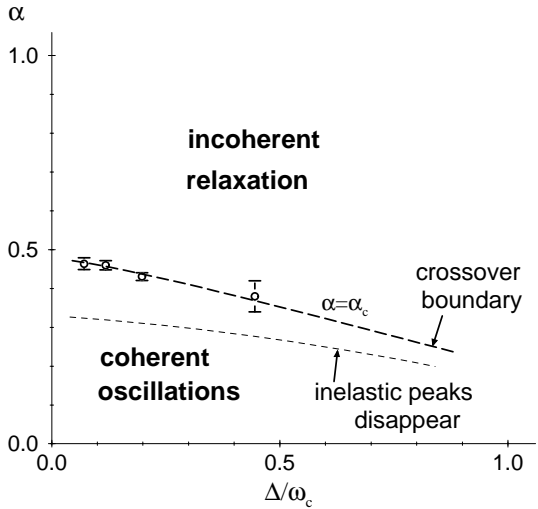}}
\caption{The phase diagram of the two-state system in the 
  $(\Delta, \alpha)$ plane. The four data points for $\alpha_c$ were
  obtained from power-law fits to $\omega_0(\alpha)$; the dashed line
  is a guide to the eye only. The values for $\Delta/\omega_c$ include
  the cutoff corrections estimated in sec.\ \ref{refShiba}. }
\label{PhaseDiagram}
\end{figure}

Our results indicate a crossover value $\alpha_c$ that is slightly
smaller than $1/2$. We have to be aware that it is in general a
difficult task to extract critical coefficients and critical parameter
values from Monte Carlo simulation data. Simulation errors increase
significantly near the transition point, and in our case the errors
get magnified by the procedure of analytical continuation to the real
axis. Nevertheless, according to our results, $\omega_0$ is strictly
zero at least for $\alpha \gtrsim 0.48$. However, for $\Delta/\omega_c
< 0.1$ the Ising correlation function falls off so fast that the
Pad\'e approximation becomes increasingly unreliable.  The
calculations for $\Delta/\omega_c = 0.05$ and $\alpha$ near its
critical value were performed on a lattice of $N = 512$ Ising spins,
but simulations with an even larger number would be necessary to
confirm the accuracy of the results. These are very difficult to
conduct due to the $N^2$ behavior of the computation time. As we shall
see in sec.\ \ref{refShiba}, cutoff corrections become significant for
small $\Delta/\omega_c$, and prevent us from approaching the limit
$\Delta/\omega_c \to 0$. With cutoff corrections and error bars taken
into account, the results presented here allow the conclusion that
$\alpha_c = 1/2$ is the correct crossover value \cite{StockburgerMak}
in the limit $\Delta/\omega_c \to 0$.

The results of the power-law fits are summarized in the following
table $(\nu$ is the ``critical exponent'' of $\omega_0$). The phase
diagram of the system is sketched in Fig.\ \ref{PhaseDiagram}.

\begin{center}
\begin{tabular}{c|c|c}
$\Delta/\omega_c$ & $\alpha_c$ & $\nu$ \\
\hline
 $0.05$ & $0.464 \pm 0.015$ & $1.11 \pm 0.11$ \\
 $0.1$ & $0.460 \pm 0.012$ & $0.91 \pm 0.12$ \\
 $0.2$ & $0.431 \pm 0.010$ & $0.63 \pm 0.09$ \\
 $0.5$ & $0.38 \pm 0.04$ & $0.47 \pm 0.31$ \\
\end{tabular}
\end{center}

\section{The biased two-state system}
\label{secBias}

\begin{figure}
  \centerline{\epsfxsize=3.4in \epsffile{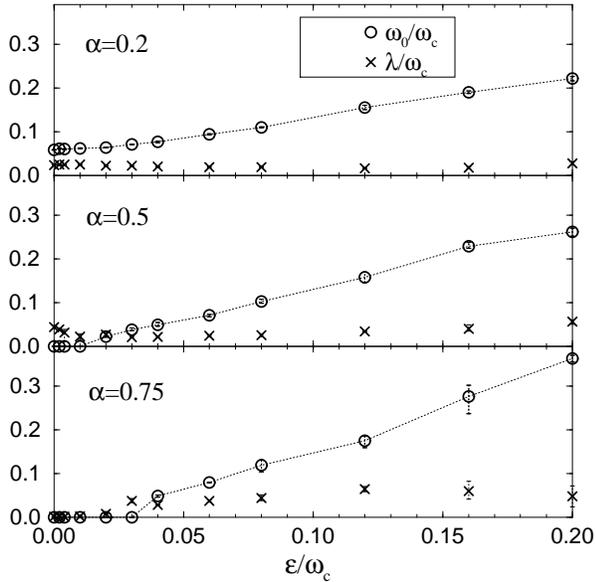}}
\caption{Tunneling frequency and damping coefficient as a function of
  the bias $\epsilon$ for $\Delta/\omega_c = 0.1$ and different values
  of $\alpha$. The biased system favors coherent relaxation. }
\label{Bias}
\end{figure}

We will now investigate the influence of a bias ($\epsilon > 0$) on
the dynamical behavior of the system, i.e. one of the states is
energetically favored. The spin-spin correlation function, as defined
above, then acquires a delta function peak at zero frequency, due to
the non-vanishing expectation value $\left< \sigma_z(\tau) \right>$.
If we instead consider the irreducible correlation function $\left<
  \sigma_z(\tau) \sigma_z(0) \right> - \left< \sigma_z \right>^2$, we
discover that its Fourier transform is again surprisingly well
described by a four-pole structure. The bias shifts the location of
the quasiparticle poles in the complex plane, but does not otherwise
change the analytic structure of the dominant part.  Fig.\ \ref{Bias}
shows the oscillation frequency $\omega_0$ and damping coefficient
$\lambda$ as a function of $\epsilon/\omega_c$, for $\Delta/\omega_c =
0.1$ and three different values of $\alpha$.


\begin{figure}
  \centerline{\epsfxsize=3.4in \epsffile{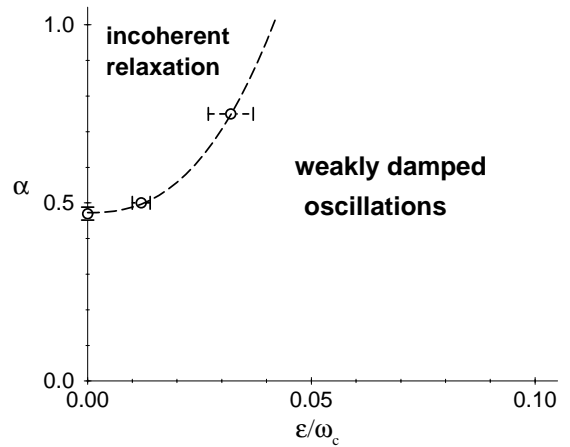}}
\caption{The phase diagram in the $(\epsilon, \alpha)$ plane, 
  for $\Delta/\omega_c = 0.1$ }
\label{Phase2}
\end{figure}

In the first graph, $\alpha = 0.2,$ the damping coefficient $\lambda$
remains largely unaffected by the bias.  As we might expect from the
undamped case $(\alpha = 0)$, where $\omega_0 = \sqrt{ \Delta^2 +
  \epsilon^2 }$, the tunneling frequency increases linearly with the
bias, in the limit of large $\epsilon$. This leads to an overall
linear increase in the $Q$ value.  These features were already observed
in Sec.\ \ref{secResults}.  The graphs for $\alpha = 0.5$ and $\alpha
= 0.75$ show the same qualitative behavior for large
$\epsilon/\omega_c$.  We see that a bias can induce a transition from
strongly damped $(\omega_0 = 0)$ to oscillatory $(\omega_0 > 0)$
behavior.  Since the system dwells in the lower-energy state for most
of the time, the interaction with the environment is reduced, and
damping is less effective in the presence of a bias. At zero
temperature and for moderate coupling to the environment ($0.5 \le
\alpha \le 1.0$), the system will always display oscillatory behavior
if the bias exceeds a critical value $\epsilon_c$
($\epsilon_c/\omega_c \simeq 0.01$ for $\Delta/\omega_c = 0.1$,
$\alpha = 0.5$, and $\epsilon_c/\omega_c \simeq 0.03$ for
$\Delta/\omega_c = 0.1$, $\alpha = 0.75$). The resulting phase diagram
is sketched in Fig.\ \ref{Phase2}.


\begin{figure}
  \centerline{\epsfxsize=3.4in \epsffile{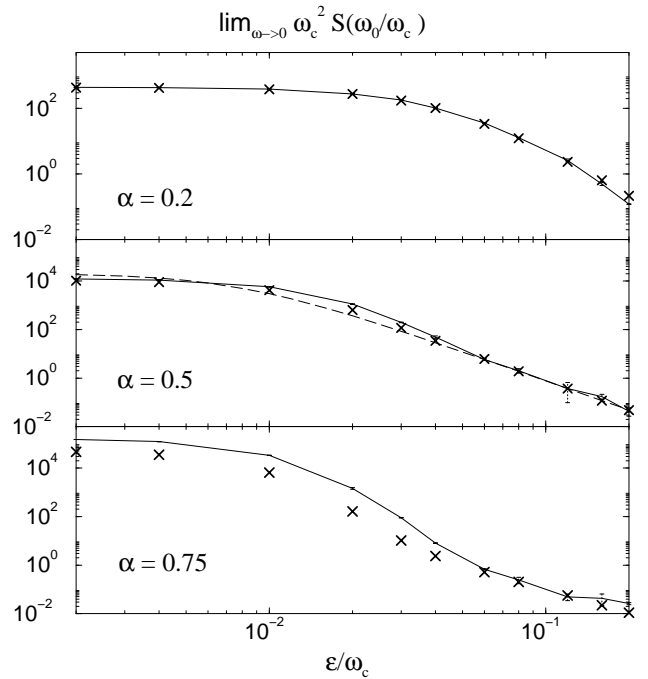}}
\caption{Zero-frequency limit of $S(\omega)$ as a function of the
  bias. The solid lines (with error bars) are our simulation results.
  The 'X' show $ S(\omega \to 0)$ as given by Shiba's relation.  The
  dashed line in the graph for $\alpha=0.5$ is the exact analytical
  expression (\ref{eqnToulouseBias}) for the Toulouse case.  }
\label{ZeroLimit}
\end{figure}

At long time scales, the dynamics are governed by the zero-frequency
behavior of the spectral function. Fig.\ \ref{ZeroLimit} shows
$\lim_{\omega \to 0} S(\omega)$ as a function of the bias. The
theoretical prediction \cite{LesSaleur2} for large $\epsilon/\omega_c$
is

\begin{equation}
  \lim_{ \omega \to 0 } \
  S \left( \omega \right) \; \sim \; \left (
      \frac{\epsilon}{\omega_c} \right)^{4 \alpha - 6}.  
\end{equation}

\nid This is in good agreement with the results shown.


\section{Shiba's relation and the Toulouse limit}
\label{refShiba}

Shiba's relation was originally proven for the Anderson model, but
later generalized to the two-state system without \cite{SassettiWeiss}
and with \cite{LesSaleur2} a bias. It connects the zero-frequency
behavior of the spectral function to the static susceptibility as:

\begin{equation}
\label{eqnShiba}
  \lim_{\omega \to 0} S(\omega) = \frac{\pi}{2} \alpha \chi_0^2,
\end{equation}

\nid where the static susceptibility

\begin{eqnarray}
\label{eqnStatSusc}
  \omega_c \chi_0 &=& \lim_{\omega \to 0} \omega_c \chi'(\omega) \nl
         &=& \sum_\tau \Bigl[ \left< \sigma_z(\tau) \sigma_z(0)
         \right> - \left< \sigma_z \right>^2 \Bigr]
\end{eqnarray}

\nid can be directly extracted from the Monte Carlo simulations on the
Ising system. Shiba's relation therefore constitutes another important
check of our numerical approach.

In the Toulouse case, $\alpha = 1/2$, the explicit form of the
spectral function is given as \cite{RMP} (note the different
normalization of the spins)

\begin{eqnarray}
\label{eqnToulouse}
  S(\omega) &=& \frac{8}{\pi \Delta_T^2} 
        \frac{1}{\left(\frac{\omega}{\Delta_T}\right)^2 + 4} 
        \biggl[ \frac{\Delta_T}{\omega} \tan^{-1} \left(
        \frac{\omega}{\Delta_T} \right) \nl
            & & + \left( \frac{\Delta_T}{\omega}
        \right)^2 \ln \left( 1 + \left( \frac{\omega}{\Delta_T}
          \right)^2 \right) \biggr],
\end{eqnarray}

\nid where

\begin{equation}
  \Delta_T = \frac{\pi}{4} \frac{\Delta^2}{\omega_c}.
\end{equation}

\begin{figure}
  \centerline{\epsfxsize=3.4in \epsffile{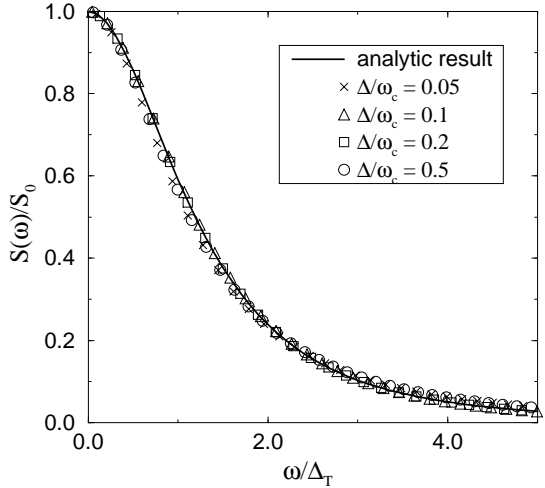}}
\caption{The spectral function in the Toulouse 
  limit $\alpha = 1/2$. The solid line shows the analytic expression,
  while the symbols represent numerical results for different values
  of $\Delta/\omega_c$. }
\label{Toulouse}
\end{figure}

\nid Figure \ref{Toulouse} compares the analytic expression to our
numerical results. The simulation data for $0.1 \le \Delta/\omega_c
\le 0.2$ reproduce the analytical result exactly. Outside this range,
the agreement is still good but no longer exact.

%
%

At zero frequency, expression (\ref{eqnToulouse}) reduces to

\begin{equation}
\label{eqnZeroToulouse}
  \lim_{\omega \to 0} S(\omega) = \frac{4}{\pi \Delta_T^2}.
\end{equation}

\nid Equations (\ref{eqnShiba}) and (\ref{eqnZeroToulouse}) allow us
to estimate the cutoff corrections to $\Delta/\omega_c$ (see section
\ref{lblSection2}) through the relation

\begin{equation}
  \frac{\Delta}{\omega_c} = \frac{4}{\pi} \left( \omega_c \chi_0
    \right)^{-1/2}.
\end{equation}

\nid The following table compares the static susceptibility, as given
by (\ref{eqnStatSusc}) and calculated from Shiba's relation, and lists
the corrected values of $\Delta/\omega_c$. Shiba's relation is
satisfied to great accuracy for larger values of $\Delta/\omega_c$,
but the agreement becomes worse for smaller values $(\Delta/\omega_c
\lesssim 0.1)$. The discrepancy in $\Delta/\omega_c$ is typically
around $10 \%$, but becomes larger for very small $\Delta/\omega_c$.

\begin{center}
\begin{tabular}{c|c|c c c|c c}
\squeezetable
$\Delta/\omega_c$ & $\omega_c \chi_0$ & $\omega_c^2 S(0)$ &
$\omega_c \chi_0$ & err. & $(\Delta/\omega_c)_{\text{corr}}$ & err. \\
  & Ising & & Shiba & Shiba & & cutoff \\
\hline
 $0.05$ & $324$ & $1.19 \times 10^5$ & $389$ & $20 \%$ & $0.071$ & $41 \%$  \\
 $0.1$ & $117$ & $1.22 \times 10^4$ & $124.6$ & $6.5 \%$ & $0.118$ & $18 \%$ \\
 $0.2$ & $41.4$ & $1.35 \times 10^3$ & $41.5$ & $0.25 \%$ & $0.198$ & $1.0 \%$ \\
 $0.5$ & $8.19$ & $53.6$ & $8.26$ & $0.87 \%$ & $0.445$ & $11.0 \%$ \\
\end{tabular}
\end{center}

\nid For the biased system, a generalization of expression
(\ref{eqnZeroToulouse}) is given by Lesage and Saleur
\cite{LesSaleur2}:

\begin{equation}
\label{eqnToulouseBias}
  \lim_{\omega \to 0} S(\omega) = \frac{4}{\pi}
  \frac{\Delta_T^2}{ \left( \epsilon^2 + \Delta_T^2 \right)^2 }.
\end{equation}

\nid Figure \ref{ZeroLimit} compares this to the numerical results. The
agreement with Shiba's relation is excellent for small $\alpha$, but
only qualitatively right for larger values of $\alpha \gtrsim 1/2$.

\section{Conclusions}

We investigated the dynamical behavior of a two-state system, coupled
to a bosonic environment with linear spectral density, in the
parameter range $0.05 \le \Delta/\omega_c \le 0.5$, $\alpha \le 0.75$
and $0 \le \epsilon/\omega_c \le 0.2$.  We found that the crossover
from oscillatory behavior to incoherent exponential decay occurs in
the vicinity of $\alpha = 1/2$, which corresponds to the Toulouse
limit of the anisotropic Kondo model.  However, technical difficulties
prevent us from obtaining precise answers in the limit of very small
$\Delta/\omega_c$: The strong nearest-neighbor interaction in the
Ising picture causes the imaginary-time correlation function
$\cal{C}(\omega)$ to fall off very fast, so that we are effectively
left with just a small number of data points on which we can base the
Pad\'e approximation.  Simulations on a larger Ising-spin system would
be necessary to yield stable results.

The results presented within the quasiparticle approximation support
the conclusion that $\alpha_c = 1/2$ is the correct crossover value in
the limit $\Delta/\omega_c \to 0$, but show that $\alpha_c < 1/2$ for
finite $\Delta/\omega_c$.

For $\alpha > 1/3$, the spectral function $S(\omega)$ does not show
any peaks on the real axis at finite frequencies, but is centered
around $\omega = 0$. We have seen that the system nevertheless
exhibits oscillatory behavior beyond that point, so that $\alpha =
1/3$ does not correspond to a crossover value.

A bias increases the oscillation frequency, in the same way as it does
for the decoupled system, and favors coherent oscillations to the
extent that a strong enough bias can induce a transition from
overdamped relaxation to weakly damped oscillatory behavior.  This is
true even for rather strong coupling $1/2 < \alpha < 1$.

\acknowledgments

I wish to thank Sudip Chakravarty for providing me with many of the
ideas that made this paper possible, and for continuous intellectual
support. This work was supported by a grant from the National Science
Foundation, Grant No. DMR 9531575


\begin{references}
\bibitem{CommentOnCt} note that our definition of the correlation
  function differs from the widely used definition $C(t) = \Ree \left<
    \sigma_z(t) \sigma_z(0) \right> = \frac{1}{2} \left< \left[
      \sigma_z(t), \sigma_z(0) \right]_+ \right>$. See the discussion 
  at the end of Sec.\ \ref{secAnalytic}.
\end{references}
\end{document}